\documentclass[structabstract]{aa} 
\usepackage[T1]{fontenc}
\usepackage{textcomp}
\usepackage[english]{babel}
\usepackage{amsfonts}
\usepackage{ae,aecompl}
\usepackage{natbib}

\usepackage{color}
\usepackage{graphicx}
\usepackage{psfrag}

\newcommand{\new}{}
\newcommand{\gras}{}
\newcommand{\last}{}
\newcommand{\be}{\begin{equation}}
\newcommand{\ee}{\end{equation}}
\newcommand{\bea}{\begin{eqnarray}}
\newcommand{\eea}{\end{eqnarray}}
\newcommand{\ie}{{\it i.e.}~}

\begin{document}

\title{Rossby Wave Instability and three-dimensional vortices in accretion disks}
\author{Heloise Meheut, Fabien Casse, Peggy Varniere, Michel Tagger}

   \author{H. Meheut
          \inst{1},
          F. Casse\inst{1},
          P. Varniere\inst{1}
          \and
          M. Tagger\inst{2,1}
          }

   \institute{AstroParticule et Cosmologie (APC), Universit\'e Paris Diderot, 10, rue A. Domon et L. Duquet 75205, Paris Cedex 13, France \\
              \email{hmeheut@apc.univ-paris7.fr}
         \and
             Laboratoire de Physique et Chimie de l'Environnement et de l'Espace (Universit\'e d'Orl\'eans/CNRS), Orl\'eans, France\\
             }


 
  \abstract
   {The formation of vortices in accretion disks is of high interest in various astrophysical contexts, in particular for planet formation or in the disks of compact objects. But despite numerous attempts it has thus far not been possible to produce strong vortices in fully three-dimensional simulations of disks.} 
   {The aim of this paper is to present the first 3D simulation of a strong vortex, established across the vertically  stratified structure of a disk by the Rossby Wave Instability.}
{Using the \gras{Versatile Advection Code (VAC), we set up a fully 3D cylindrical stratified disk potentially prone to the Rossby Wave Instability.}}
   {The simulation confirms the basic expectations obtained from previous 2D analytic and numerical works. \gras{The simulation exhibits} a strong vortex that grows rapidly and saturates at a finite amplitude. On the other hand the third dimension shows unexpected additional behaviours that could be of strong importance in the astrophysical roles that such vortices can play.}
   {}

   \authorrunning{Meheut et al.}
   
   \titlerunning{RWI and 3D vortices in accretion disks}

   \maketitle
%


\section{Introduction}

Vortices have been actively searched for in accretion disk theory and numerical simulations because of their  multiple astrophysical interests. In particular, in the case of protoplanetary disks, the presence of vortices has been invoked \citep{BAR95} to alleviate the problem of the too long time scale needed for the growth of grains to planetesimals. The streaming of the gas inside a vortex could speed up this growth by concentrating dust grains in its centre, by what is sometimes called the \lq tea leaf\rq~effect whereby tea leaves accumulate in the center of a cup when it is stirred. This mechanism has been studied numerically \citep{JOH04,LYR08, LYR09} but only in the 2D case of an infinitely thin disk.\gras{ We note that an alternative mechanism for trapping the dust grains based on high pressure region has been proposed by \cite{JOH07}.}
 
%

On the other hand, despite multiple attempts to generate realistic vortices in fully cylindrical 3D numerical simulations of disks, even the most advanced ones \citep{BAR05} have only found off-midplane vortices. This may be explained \citep{TAG01} by the fact that due to both differential rotation and differential vorticity (which is absent in the Shearing Box model used in many simulations) vortices are very rapidly sheared away and could survive only by being continuously re-generated.

In this paper we present the first 3D numerical simulations of the Rossby Wave Instability (RWI), which is precisely a mechanism to do this: in certain conditions this instability generates Rossby vortices that grow exponentially with time.\gras{ The existence of this instability in protoplanetary disks was first mentioned in \cite{RWI1} and} it has been suggested later \citep{VAR06} that at the edges of the \lq Dead Zone\rq~of protoplanetary disks, where the gas is not ionized and thus should not be turbulent, the conditions are such that the RWI \gras{can grow}; it should thus both help accretion to proceed across the Dead Zone, and help planets to grow there. The simulations we present confirm the basic properties of the instability, but they show new features in its 3D structure, whose effects on accretion and the growth of planetesimals will need to be understood.

The paper is organized as follows: after recalling the basic properties of the RWI we will present the physical setup, then the numerical one. Section 4 will present the results, and section 5 a discussion and conclusion.
%
\section{The Rossby Wave Instability}\label{sec:Rossby}
%

   \begin{figure}
   \centering
    \includegraphics[width=8.8cm]{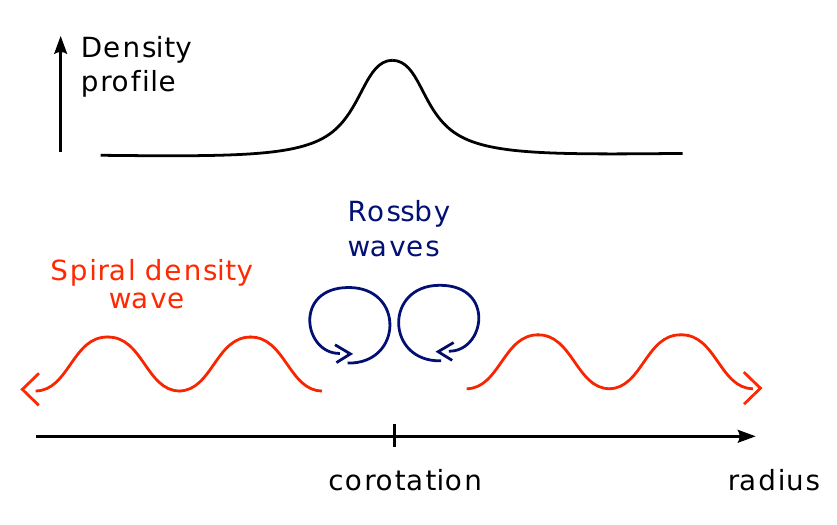}
      \caption{Schematic view of the Rossby Wave Instability. Rossby waves are generated in the region of the density extremum, and density waves are emitted away from this region due to differential vorticity, which couples these two waves. The Rossby waves have their corotation radius at the density maximum}
         \label{FigSchema}     
   \end{figure}
%

The RWI has been found and discussed in different contexts of differentially rotating disks. A discussion of this history has been given by \cite{VAR06}, and we will not repeat it here. It can exist in galactic disks
\citep{LOV78, SEL91}, as well as in accretion disks (\cite{PAP85, RWI1}; see also \cite{RWI2,RWI3}). It can be seen as the form that the Kelvin-Helmoltz instability takes in differentially rotating disks, and it has a similar instability criterion: the existence of an extremum of a quantity $\mathcal{L}$ related to the vorticity of the equilibrium flow. In an unmagnetized thin disk this quantity can be written as \citep{RWI2}:
\be
\mathcal{L}=\frac{\Sigma \Omega}{2\kappa ^2}\frac{p}{\Sigma ^ \gamma}=\frac{\Sigma}{2(\vec \nabla \times \vec v)_z}\frac{p}{\Sigma ^ \gamma}
\ee
where $\Sigma$ is the surface mass density, $\vec v$ is the velocity of the fluid, $\gamma$ the adiabatic index, $\Omega$ the rotation frequency and $\kappa^2=4\Omega+2\Omega \Omega '$ the epicyclic frequency squared (so that $\kappa^2/2\Omega$ is the vorticity). \gras{Here the prime notes a radial derivative. }

Two  possibilities that can result in an extremum in $\mathcal{L}$ have been investigated: the first one occurs near  the marginally stable orbit around a compact object, where relativistic effects create a maximum of $\kappa^2/2\Omega$. The growth rate of the RWI is strongly increased by a poloidal magnetic field threading the disk \citep{TAV06}, but decreased by a toroidal one \citep{YU09}. This  has been proposed as an explanation for the high frequency quasi-periodic oscillation (QPO) of microquasars \citep{TAV06, LAI09, TSA09}. 

The second possibility is to have an extremum of the surface density. The model of \cite{VAR06} for protoplanetary disks relies on the fact that extrema of $\Sigma$ should occur at the edges of the Dead Zone of these disks, so that the RWI should be unstable there. \cite{TAG06} and \cite{FAL07} have also used the RWI in an explanation for the quasi-periodicity that may have been observed during the flares of SgrA*. 

The RWI  has been studied both analytically \citep{RWI1,RWI2} and numerically \citep{RWI3, VAR06}. It is formed by Rossby waves trapped in the extremum of $\mathcal L$, as shown in figure~\ref{FigSchema}. \gras{In the $\beta$-plane approximation differential rotation is neglected and the dispersion relation of Rossby waves is given by 
\be
\omega=\frac{m\alpha}{k^2+m^2}
\ee
where $k$ is the adimensional radial wavenumber, $m$ the azimuthal wavenumber, and $\alpha$ the vorticity gradient. In an accretion disk,} differential vorticity and differential rotation couple them to spiral waves, emitted on both sides of the extremum region. \gras{The wave dispersion relation calculated from the equations obtained by \cite{RWI1} and \cite{TAG01} is
\be
\tilde\omega=\frac{c_s^2m\alpha}{c_s^2(k^2+m^2)+r^2\kappa^2}
\label{EqDisp}
\ee
where $\tilde\omega=\omega-m\Omega$, $c_s$ is the sound speed and $r$ the radius.}

Rossby waves have their corotation radius (where their phase velocity equals the rotation velocity of the gas) at that extremum.  The standing wave pattern they form appears as a vortex located in the region of the extremum, and grows exponentially. The waves have a positive flux of energy and angular momentum beyond that radius, and a negative flux inside it, so that (as will be checked in the simulation) the pattern can grow as it causes the gas inside corotation to lose angular momentum and accrete, while the gas beyond corotation gains that angular momentum and moves outward. 

As a result the instability tends to destroy the extremum of $\mathcal L$, as seen for instance in \cite{TAG06}. In the context of the mechanism proposed by \cite{VAR06} for protoplanetary disks, we expect the instability to saturate at the amplitude where this is balanced by the continuous regeneration of the extremum of density by the gas accreting from the outer disk and piling up at the edge of the Dead Zone (and oppositely at the inner edge of the Dead Zone). 

However these studies have all been done in the approximation of an infinitely thin disk, because of the complexity of a full 3D analytical study, or of the numerical resources needed for 3D simulations. Such a study is however highly desirable, both to consider the full complexity of the gas (and grains) flow, and to validate the thin disk approximation: this approximation was introduced (and its limits defined) by \cite{GOL65-1,GOL65-2}, but this does not apply as such to Rossby perturbations. We will see farther in this paper that indeed 3D effects bring in significant and potentially important qualitative differences.

%
\section{Gaseous accretion disk}
\subsection{Hydrodynamics equations}

We work in cylindrical coordinates $(r, \phi, z)$ with the 3D Euler equation:
\bea
\partial_t \rho +\vec  \nabla . (\vec v \rho)&=&0\\
\partial_t (\rho \vec v) + \nabla . (\vec v \rho \vec v)+\vec \nabla p&=& -\rho \vec \nabla \Phi_G\label{momentum}
\eea
where $\rho$ is the mass density of the fluid, and $\vec v$ its velocity, and $p$ the pressure. $\Phi_G={-GM_*}/{(r^2+z^2)^{1/2}}$ is the gravity potential of the central object with G the gravitational constant and $M_*$ the mass of the central object. 
We consider a barotropic flow, \ie the entropy $S$ is constant in the entire system:
\be
p=S\rho^\gamma\nonumber
\ee
with the adiabatic index $\gamma=5/3$. The sound speed is given by $c_s^2=\gamma p/\rho=S\gamma \rho^{\gamma-1}$ and the temperature by $T=p/\rho=S\rho ^{\gamma -1}$. 
\subsection{Initial conditions}\label{sec:init}
%

   \begin{figure}
   \centering
     \includegraphics[width=9.6cm]{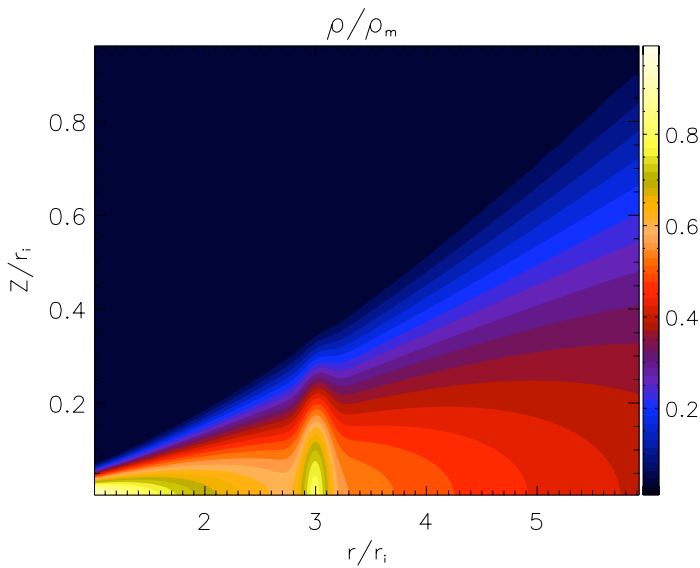}
      \caption{Isocontours of the initial density, with a bump at $r_B=3r_i$. The density is normalized to the density in the midplane at the inner edge.}
         \label{FigCI}
   \end{figure}
%

   \begin{figure}
   \centering
     \includegraphics[width=9.4cm]{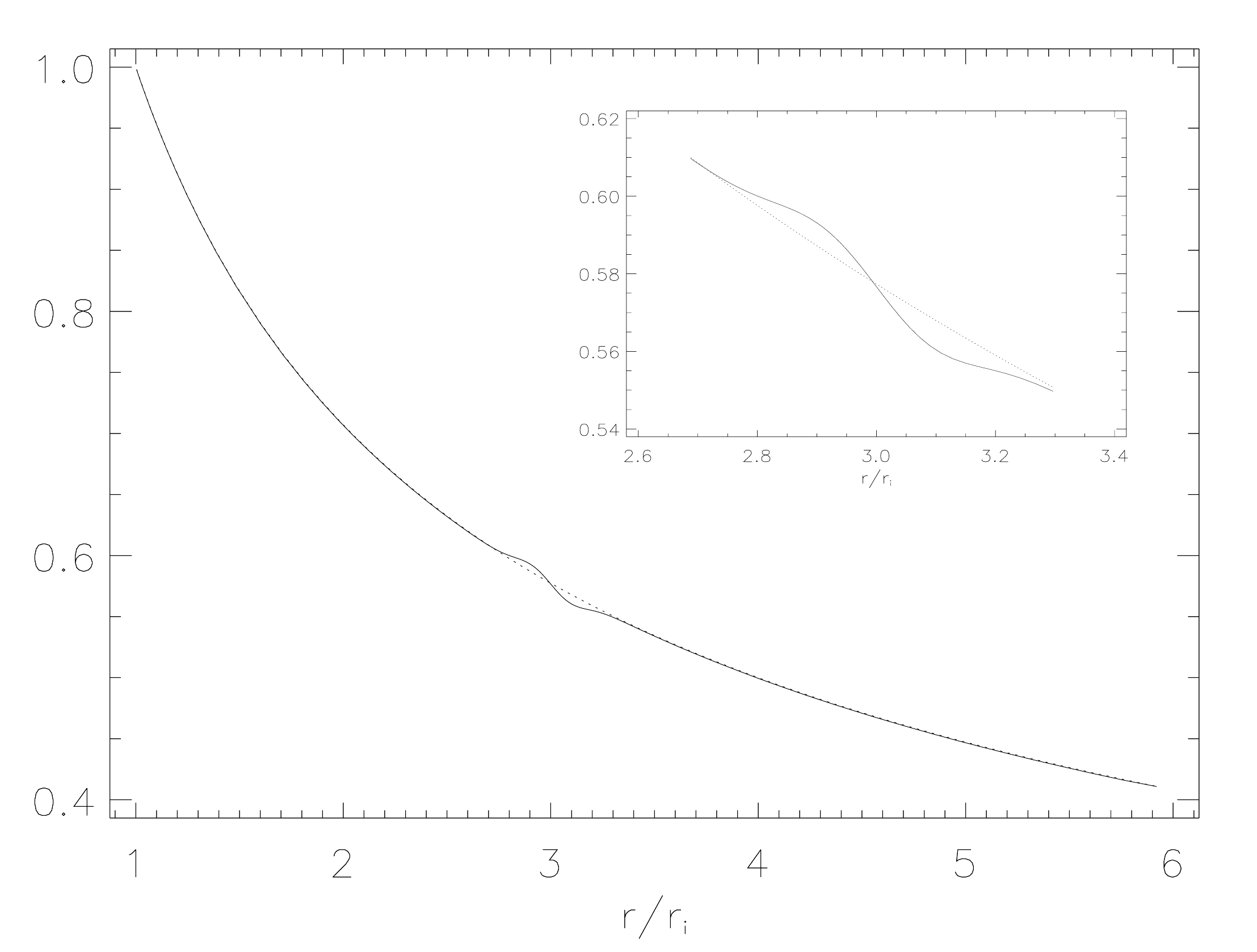}
      \caption{\gras{The initial azimuthal velocity of the midplane gas (continuous line) is compared to keplerian rotation (dots). A zoom on the bump region is shown in the upper right corner.
      }}
         \label{FigCIvk}
   \end{figure}
%

We choose as initial equilibrium a density profile decreasing radially as $r^{-1/2}$, to which is added a density bump:
 \be
 \frac{\rho^{ini}(r,z=0)}{\rho_m}=\Big(1+\chi \exp{\displaystyle \frac{(r/r_i-r_B/r_i)^2}{2\sigma ^2}}\Big)\frac{1}{\sqrt{r/r_{i}}}\\
 \ee
%
%

   \begin{figure}
   \centering
    \includegraphics[width=8.8cm]{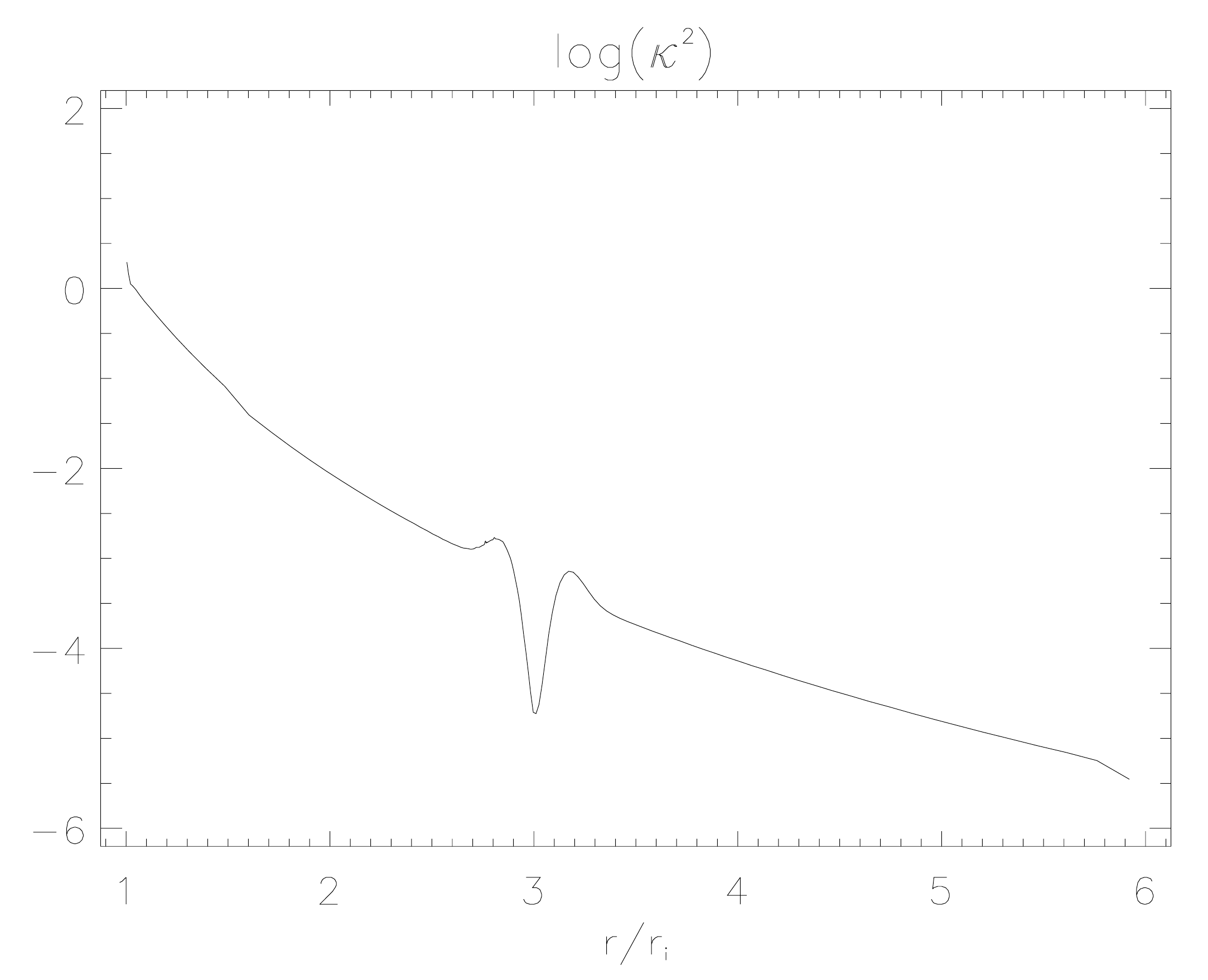}
      \caption{\last{Plot of the logarithm of the square of the epicyclic frequency $\kappa^2$ (averaged over z) as a function of the radius.}}
         \label{FigKappa}
   \end{figure}

The density is normalized by $\rho_m$, the midplane density at the inner boundary of the simulation $r_{i}$, and $r_B$ is the position of the bump.
For the parameters  $\chi$ and $\sigma$, which control the amplitude and the width of the density bump, we take the values $\chi=0.4$ and $\sigma=0.1$ respectively \gras{that gives a bump similar to the one obtained by \cite{VAR06} in their simulation of the dead zone}. These values allow the instability to grow relatively fast, decreasing the numerical load, while the effect of the pressure gradient on the equilibrium rotation velocity leaves us a margin with respect to the Rayleigh criterion $\kappa^{2}>0$ \gras{(see figure \ref{FigKappa})}. 

The bump is centered at $r_{B}=3 r_i$, far enough from the inner edge of the disk to avoid a strong effect of the boundary condition there. The vertical density profile is initially at hydrostatic equilibrium, giving an aspect ratio of the order of $10^{-1}$. 
\bea
\rho^{ini}(r,z)=\rho(r,z=0)\Big[1-\frac{\gamma-1}{\gamma S \rho(r,z=0)^{\gamma-1}}\\ \nonumber
\Big(\frac{r_i}{r}-\frac{r_i}{\sqrt{r^2+z^2}}\Big)\Big]^{\frac{1}{\gamma-1}}
\eea
with \gras{and $S=10^{-3}r_i^2\Omega_K(r_i)^2\rho_m^{1-\gamma}$}.
Finally, we use initially a \lq floor\rq~  density $\rho_{min}=10^{-2}\rho_m$ in order to avoid getting too low densities in the corona. Figure \ref{FigCI} shows the resulting isodensity contours in a vertical cut of the disk.
%


The initial velocity field is purely toroidal, $v_r^{ini}=v_z^{ini}=0$ and $v_{\phi}^{ini}$ has been chosen for the disk to be in radial equilibrium in a Newtonian potential:
 \be
  v_{\phi}^{ini}=\sqrt{ \frac{ r^2GM_* r_i}{(r^2+z^2)^{3/2}}+\frac{r\rho_m}{\rho r_i}\nabla_r p}
 \ee
\gras{Figure \ref{FigCIvk} shows the deviation of the disk azimuthal velocity from a pure keplerian rotation. The pressure term in the last equation induces only a weak variation.}
\section{Numerical setup}

\subsection{Numerical code and scheme}

The numerical simulations presented here use the Versatile Advection Code (VAC) developped by \citet{TOT96}. In the version we use the code solves the 3D hydrodynamics equations for an isentropic flow. We use VAC with the total variation diminishing monotonic upstream-centered scheme for conservation laws (TVD-MUSCL), a Roe Riemann solver, a Hancock predictor step and a Woodward limiter \citep{COL84}. The TVD-MUSCL scheme detailed in \citet{TOO96} is one of the less dissipative schemes included in the VAC code, which is useful in order to observe the full development of the instability and properly characterize its saturation.
%
\subsection{Grid and numerical resolution}

The grid is cylindrical ($r$,$\phi$,$z$) with a resolution of $154\times68\times68$, including 2 ghosts cells at each boundary in order to impose the boundary conditions. The grid is non uniform \gras{with a streching factor of 20 in the radial direction, and 10 in the vertical one.
}
This achieves a higher resolution in the regions of physical or numerical interest. 
 The mesh extends from ($r_i$,0,0) to ($6r_i$, $r_i$, $2\pi$) where $r$ and $z$ have been normalized by the disk inner edge radius $r_i$. The radial and vertical extensions are large enough for the different waves of the RWI to develop without unwanted effects from the boundary conditions. 
%

The minimum value of the mass density is fixed to $10^{-6}\rho_m$. 
With the initial conditions presented before this configuration gives a bump with approximatively 60 cells radially and 40 vertically. 
%
\subsection{Boundary conditions}

The boundary conditions are imposed using the two ghost cells added at each boundary of the mesh, allowing to compute the derivatives in all the cells. Since the simulation is restricted to a small part of the disk, we have chosen continuous boundary conditions in the radial direction, meaning that the values of the physical quantities are copied from the first cell of the computational domain in the ghost zones. This boundary condition is thus  partially transparent. In the azimuthal direction the boundary condition is periodic. Only the upper part of the disk is simulated since the vortices we are interested in are even in $z$. In section \ref{velocity} we discuss a test run without this condition, which confirms the validity of this choice. 

   \begin{figure}
   \centering
    \includegraphics[width=8.8cm]{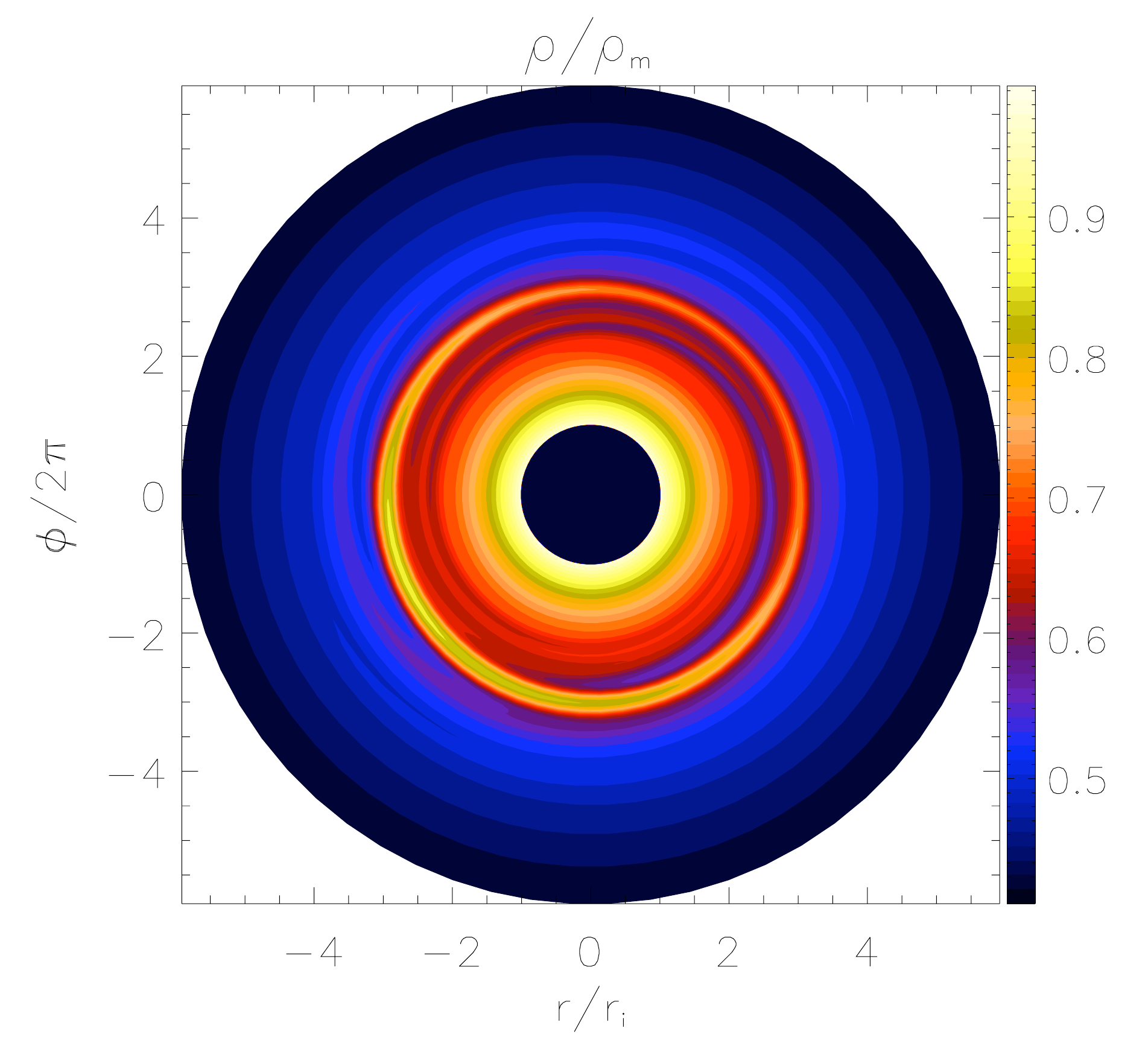}
      \caption{Plot of the normalized density in the midplane of the disk at $t=176/\Omega_K(r_i)$. One can identify the position of the crescent-shaped vortex at the bump radius ($r=3r_i$) and the spiral arms on both side of the bump region.}
         \label{fig:polaire}
   \end{figure}
 %
 \subsection{Disk equilibrium and stability}\label{equilibrium}

As mentioned in section \ref{sec:init} we start the simulation with a low but finite density in the corona. On the other hand, we need to start from a true disk equilibrium in order to avoid a rapid relaxation that would dominate the evolution of the system. The main difficulty is  the constant density we choose to start from in the corona, and the transition to  the power law vertical profile in the disk; this introduces a jump in the vertical derivative of the density. We thus apply to this density profile a smoothing over $n_{smooth}$ vertical grid points (with $n_{smooth}=4$ here) in order to enforce a smooth transition. All this means that we are not at exact hydrostatic equilibrium in the gravity field. A different but related difficulty arises from the discretization of the grid \gras{and the use of slope limiters}, which introduces residual spurious forces \gras{for each cell but a zero mean value on the whole grid}. Although they are small (at the level of roundoff error) they are systematic (\ie always act in the same direction) and become important in low-density regions, especially at the disk-corona transition. 

We thus choose to slightly modify the gravitational potential, so that the initial pressure profile given in section \ref{sec:init} is really at equilibrium. We apply the following procedure: \gras{a first time step is done to calculate the difference between the physical and numerical equilibrium. This allows us to} compute the source term $\rho \vec {\nabla}\Phi_G$ in the Euler equations \gras{that enforces numerical equilibrium}: this source term should be equal to the gravity term if we were in exact equilibrium, and the difference arises from these spurious forces. We enforce equilibrium by subsequently considering this \gras{constant} modified source term. We will a posteriori check that this does not substantially affect our result. 
\new{Moreover we have checked that with the whole simulation box inside the disk (without corona) this modified source term is not needed, but  wave reflection at the upper boundary makes it impossible to study properly the physics of the instability.}
%

\gras{After several time steps} we add to these initial conditions low level perturbations so as to provide seeds for the unstable modes that can develop if the disk is unstable. We find more convenient to do this with perturbations of low azimuthal wavenumber, since (as confirmed in the simulation) they are the only ones expected to be unstable. We thus apply an initial radial velocity perturbation:
\bea
v_{rp}=v_r+\epsilon \sin\big(2\pi(r-1.2)/.8\big)[\sin(\phi)+\sin(2\phi)\\
	+\sin(3\phi)+\sin(5\phi)]\nonumber
\eea 
with the amplitude of the perturbation \gras{$\epsilon$ of the order of $10^{-7}$}.

\section{Results}

In this section we describe the results obtained with this simulation. We first describe the general properties of the instability, which allow us to identify it and to study its structure and evolution. We then focus on the velocity stream to understand the flow pattern induced by the instability in the disk.
\subsection{General study of the instability}

%
%
   \begin{figure*}[!t]
   \label{FigDivCurl}
   \centering
 	 \includegraphics[width=7.2cm]{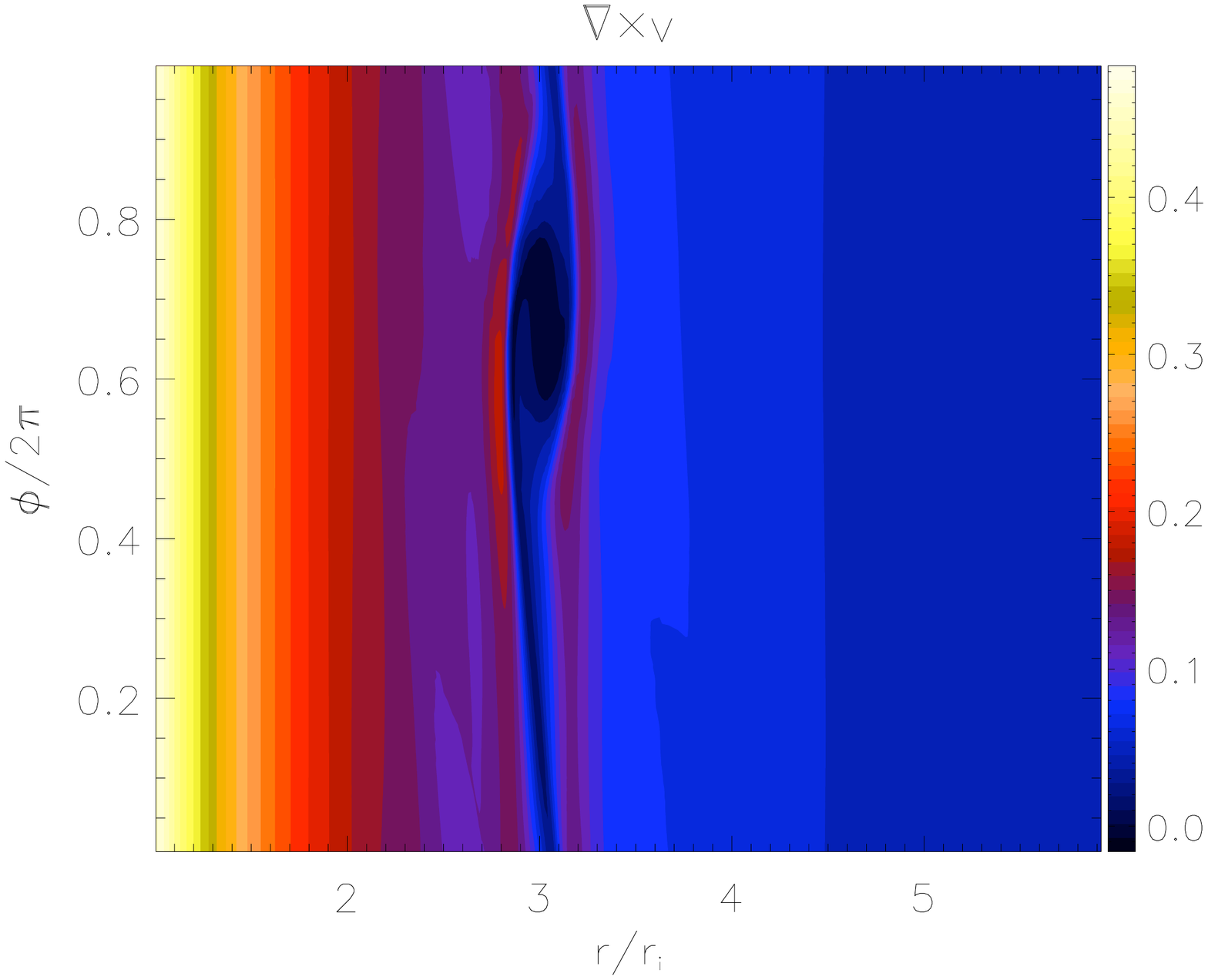}
    	 \includegraphics[width=7.2cm]{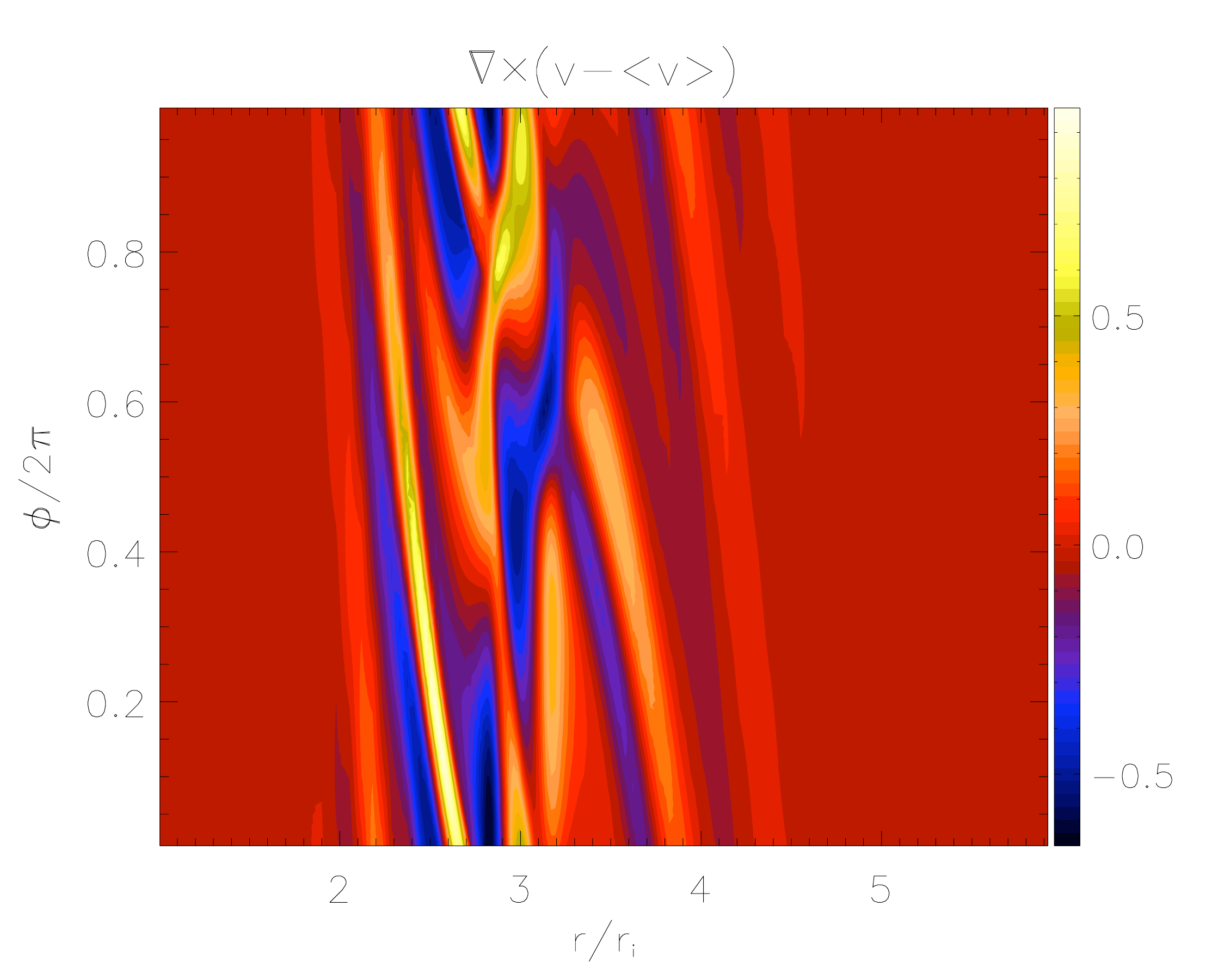}
     \includegraphics[width=7.2cm]{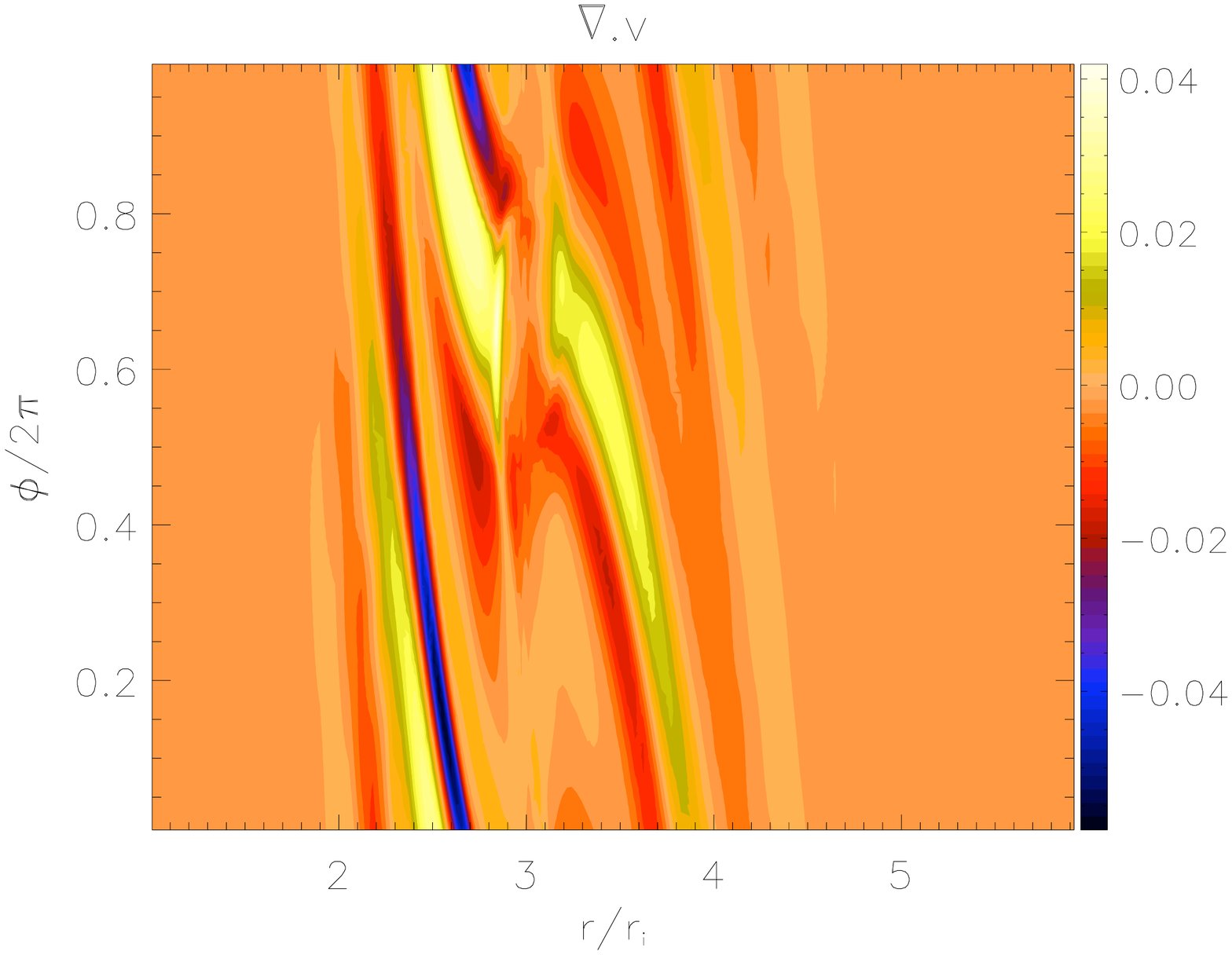}
     \includegraphics[width=7.2cm]{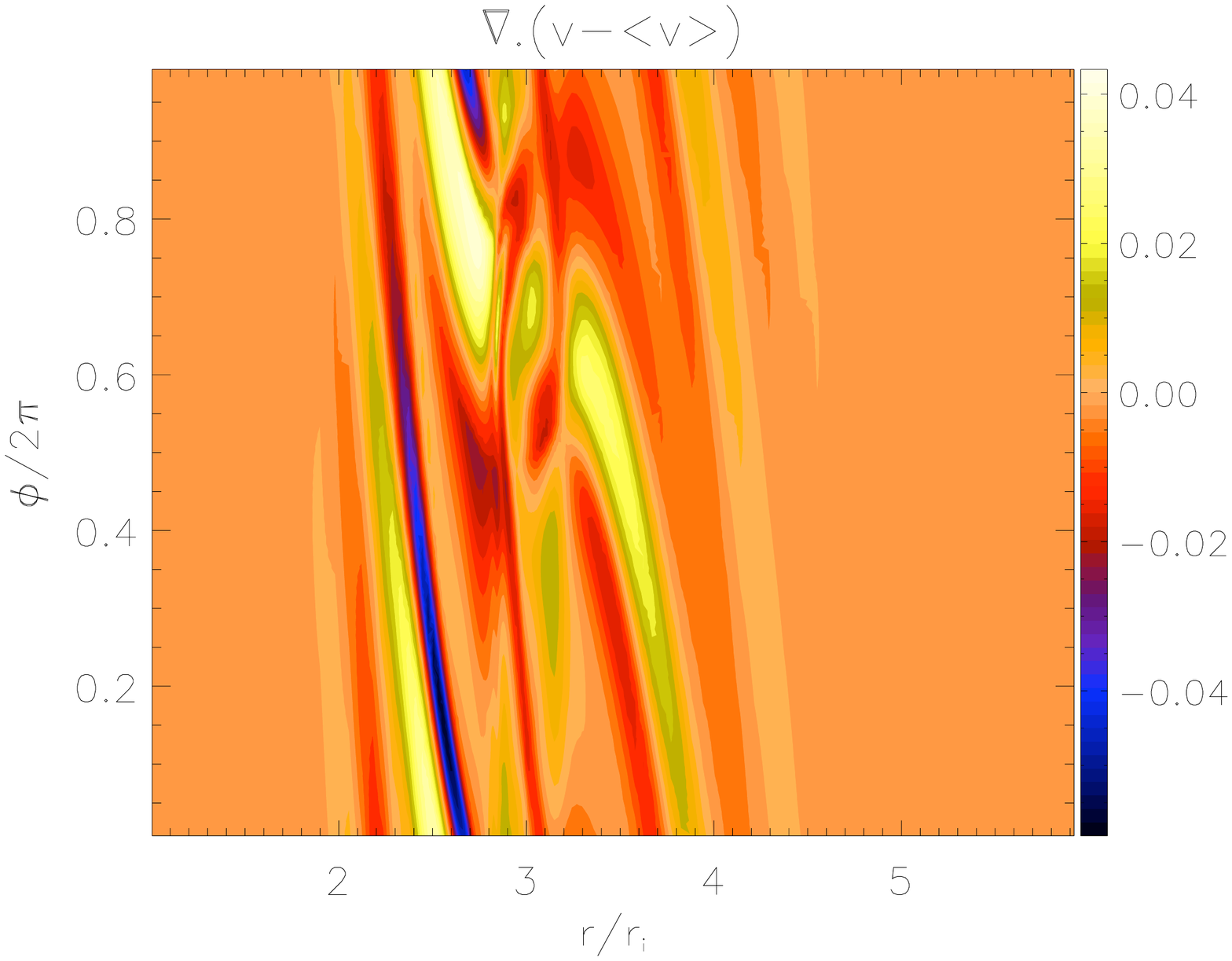}
     
\caption{
The compressional and rotational parts of the total flow  (left) and its non-axisymmetric part  (right) at $t=176/\Omega_K(r_i)$, showing the two waves that form the instability. \emph{Top}: the amplitude of $(\vec \nabla \times \vec v)_z$ traces the vorticity, \ie the Rossby vortex centered at the extremum of density. \emph{Bottom}: the amplitude of $\vec \nabla .\vec v$ traces the compressional waves (spiral density waves) that develop away from that region. In these figures, the whole disk is represented with the azimuthal coordinate on the vertical axis and the radius on the horizontal one. Note that, 
although the top-right panel shows both a cyclonic and an anticyclonic vortex (as expected for an $m=1$ mode), once combined with the bulk flow only the anticyclonic vortex appears in the top-left panel.}
\label{FigDivCurl}
\end{figure*}
Since the instability has only been studied in 2D, we will first base the discussion on what we obtain in the midplane of the 3D simulation. As expected the simulation shows after a few rotation times a coherent perturbed pattern forming in the region of the density bump. Figure \ref{fig:polaire} shows the normalized density in the midplane.  The non-axisymmetric pattern corresponds to the RWI, extending spiral arms on both sides of the bump.

As explained in section \ref{sec:Rossby}, the structure of the instability is formed by Rossby waves on both sides of the extremum of $\mathcal L$ (here, the density bump at $r=3r_i$), resulting in a standing vortex there, and spiral density waves emitted on both sides of that region. We can visualize this by plotting the curl and divergence of the flow, showing respectively its rotational (Rossby waves) and compressional (density waves) parts. Figure \ref{FigDivCurl} shows that the expected patterns do indeed appear.

\begin{figure}
   \centering
   \includegraphics[width=8.8cm]{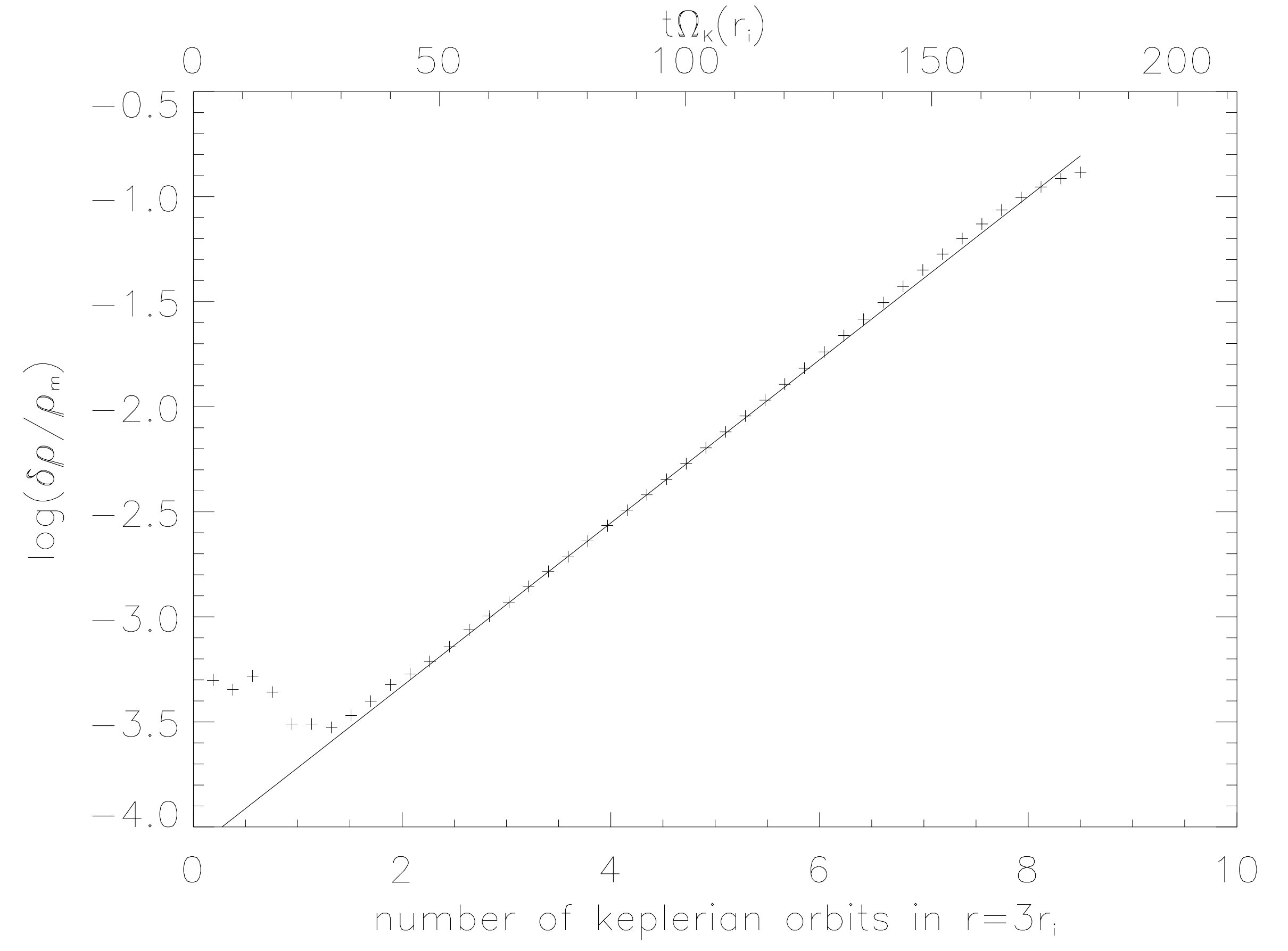}
      \caption{Instability growth: the \gras{maximum value of the} non axisymmetric part of the density is plotted as a function of time, with a logarithmic scale for the vertical axis. \gras{We have here added a second time axis (lower) that shows the number of keplerian orbit at the bump radius in order to estimate the growth rate in the same units as \cite{RWI2}.} The equation of the continuous line is \new{$-4.1+0.39t\Omega_K(r_i) / \Omega_K(3r_i)$}. The graph thus shows the linear \new{stage} and the saturation after about 10 keplerian times at $r=3r_i$.}
         \label{FigGrowth}
   \end{figure}

The different stages of the development of the instability can be seen on figure \ref{FigGrowth}, which presents on a logarithmic scale the amplitude of the density perturbation as a function of time. After a short stage where the non pertinent perturbations decrease, the unstable mode enters the linear stage, \ie the perturbation grows exponentially. \gras{ We find a growth rate of $.39\Omega_K(r=r_i)/\Omega_K(r=3r_i)$. This is comparable with the growth rates obtained from 2D theory \citep{RWI2} or numerical simulation \citep{TAG06}. The different physical and numerical setups (including the equation of state and the boundary conditions) prevent a more detailed comparison. The local dispersion equation (eq. \ref{EqDisp}) does not allow to estimate the growth rate of the global modes. This would be possible anyway only with a fully 3D linear computation, which would require a very important effort. The perfect exponential growth we obtain in the present non-linear simulation over four decades in amplitude, its coherence with the 2D results, and its independence on initial conditions show that we do capture properly here the linear phase of the instability.}
This stage lasts about 10 keplerian orbits at the bump radius, and finally the instability reaches saturation. The entire progression of the instability is thus captured in this simulation. 

Figure \ref{Figt0t280} shows that the saturation is not due to the flattening of the initial density bump and thus of the extremum of $\mathcal L$, as in the 2D simulations of \cite{TAG06}, \new{so that non-linearities must be suspected. In order to check this we have run a new simulation, taking as initial radial density profile the average one at the end of the present simulation (but again with only weak velocity perturbation, as at its startup), and observed the RWI growing again.} 

\new{The azimuthal density profile at the bump radius is shown in figure 10. It reveals a departure from a pure sinusoid, corresponding to the presence of $m > 1$ modes, whose growth is shown in figure 7. Since
they do not show the exponential growth of the $m = 1$ but they show modes that are
not present in the initial perturbation, we analyze them as non-linearly generated
harmonics of that mode. Figure 8 shows that their effect remains relatively weak, so that
the saturation of the mode cannot be ascribed to them.}

\new{On the other hand, as shown in figure \ref{FigCirculation}, small-scale structures appear in the vertical flow, and become stronger when the mode approaches saturation. They cascade to the smallest scale of the simulation, making us
believe that dissipation in these structures is the main cause of the saturation we observe. Assessing their role will require both improved numerics, to deal with these structures, and physical understanding since one can expect that vertically propagating sound waves generated within the disk will be subject to wavebreaking when they reach the low-density, low-sound speed corona. This will be considered in future work.}

%
   \begin{figure*}
   \centering
   \includegraphics[width=18cm]{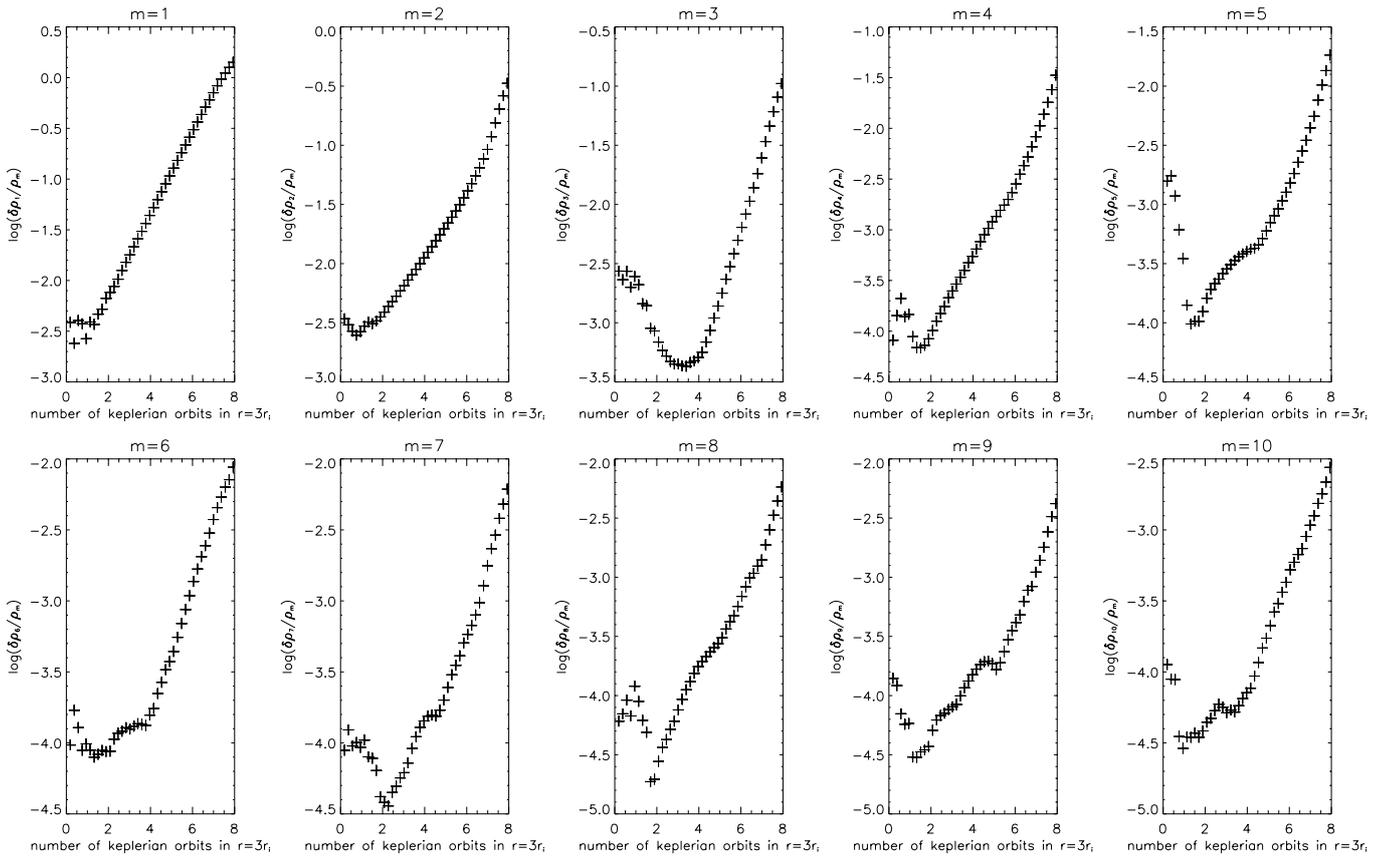}
      \caption{\gras{The evolution of the amplitude of the ten first azimuthal modes in the same coordinates as figure \ref{FigGrowth}. \new{The amplitude of each mode is calculated with an azimuthal Fourier transform, $\rho_i$ being the amplitude of the $i^{th}$ mode. The development of modes that were not present in the initial perturbation proves the existence of non-linearities.} }}
         \label{FigModesGrowth}
   \end{figure*}
%
   \begin{figure}
   \centering
  \includegraphics[width=8.8cm]{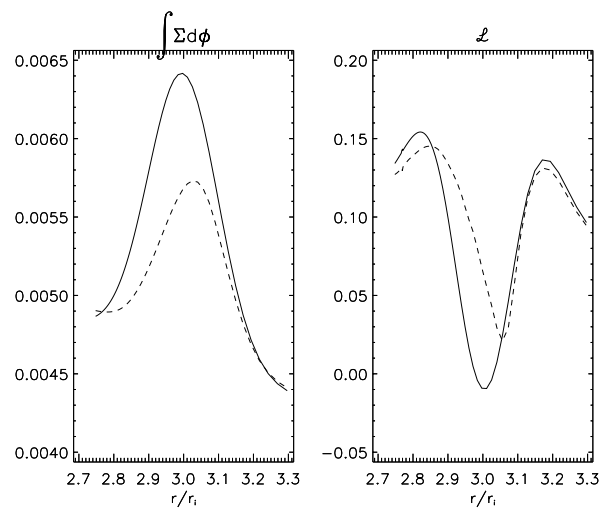}
      \caption{Comparison between the density integrated over $z$ and $\phi$ (left) and of the RWI criterion $\mathcal L$ (right) at $t=0$ in solid line and $t=176/\Omega_K(r_i)$ in dashed line. One can see that the extremum of $\mathcal L$ at $r=3r_i$ has decreased but not disappeared when the instability saturates.}
         \label{Figt0t280}
   \end{figure}
%
   \begin{figure}
   \centering
  \includegraphics[width=8.8cm]{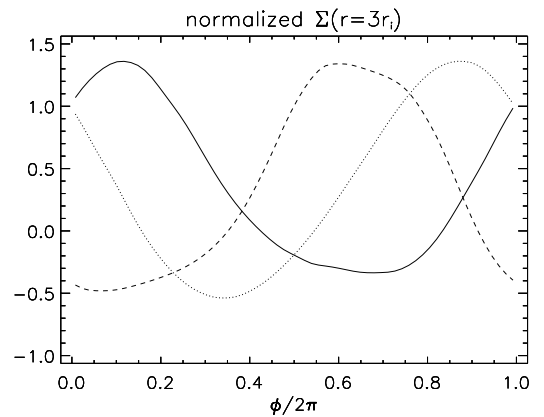}
      \caption{\new{The azimuthal surface density profile at $r=3r_i$ at different times with the same normalization. The line is at $t=32/\Omega_K(r_i)$, the departure from the sinusoid comes from the initial perturbations that is a sum of sinus. The dots is at $t=104/\Omega_K(r_i)$, the profil is closer form a sinusoid, the instability is in the linear phase, the nonpertinent modes (m>1) are negligeable. The dashed line is at $t=176/\Omega_K(r_i)$, the departure from a sinusoid \gras{is believed to be a direct consequence of non-linearities.}}}
         \label{FigSinus}
   \end{figure}
\begin{figure}
   \centering
   \includegraphics[width=8.8cm]{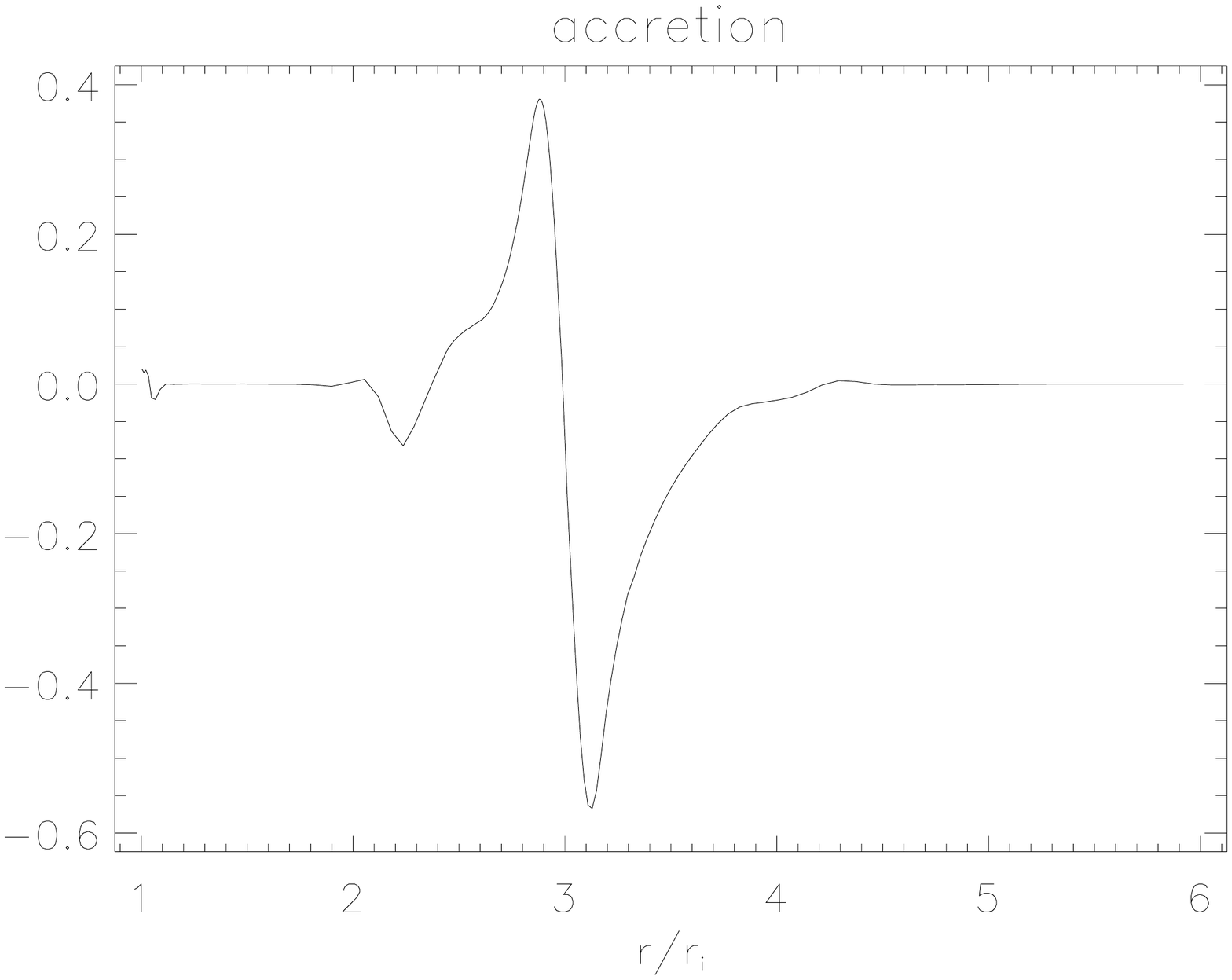}
      \caption{The accretion rate defined as $- \int r\rho v_r dz d\phi / \int r\rho_m c_s dz d\phi$ as a function of radius, at the end of the linear phase ($t=176/\Omega_K(r_i)$)}
         \label{FigAccretion}
   \end{figure}

%
%
The flattening of the bump \new{(figure \ref{Figt0t280})} is explained by the accretion rate, shown on figure \ref{FigAccretion}: it is positive inward from the density extremum and negative beyond it. This is expected from a mode with its corotation at that radius which grows by exchanging angular momentum between the inner and outer region. As in 2D simulations, the transition between positive and negative accretion is so sharp that it is actually the best diagnostic of the corotation radius and thus of the mode frequency.
%
   \begin{figure*}
   \centering
 \includegraphics[width=16cm]{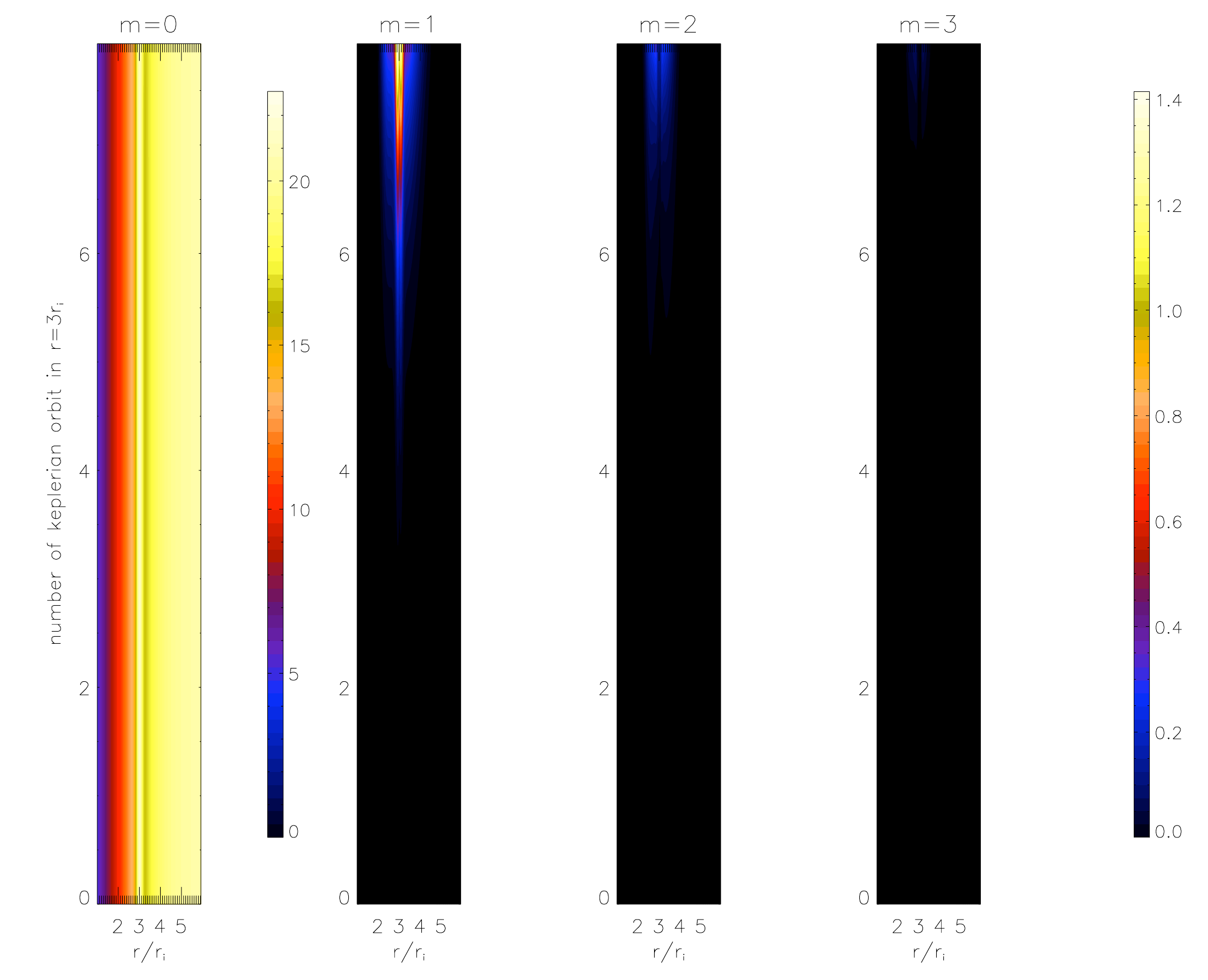}
      \caption{Temporal evolutions of the first four azimuthal modes. A different color bar is used for m=0 which is approximately ten times stronger than the others.}
         \label{FigModes}
   \end{figure*}

Finally, considering the possibility of other unstable modes, we show in figure  \ref{FigModes} the time evolution of the azimuthal Fourier components, $m=0$ to $3$. Although perturbations do appear at $m=2$ and $3$, the fact that they evolve together with the $m=1$ indicates that they are probably harmonics of that mode and not independently growing ones. The most linearly unstable $m$ depends on the local conditions. We have run 2D simulations with the same density profile and find in that case a dominant $m=2$.  We will see below that the 3D simulation does show new features in the structure of the mode, which seem to contribute to its linear growth. They might be at the origin of this difference.
\subsection{Study of the velocity stream}\label{velocity}

   \begin{figure*}
   \centering
   \includegraphics[width=8.8cm]{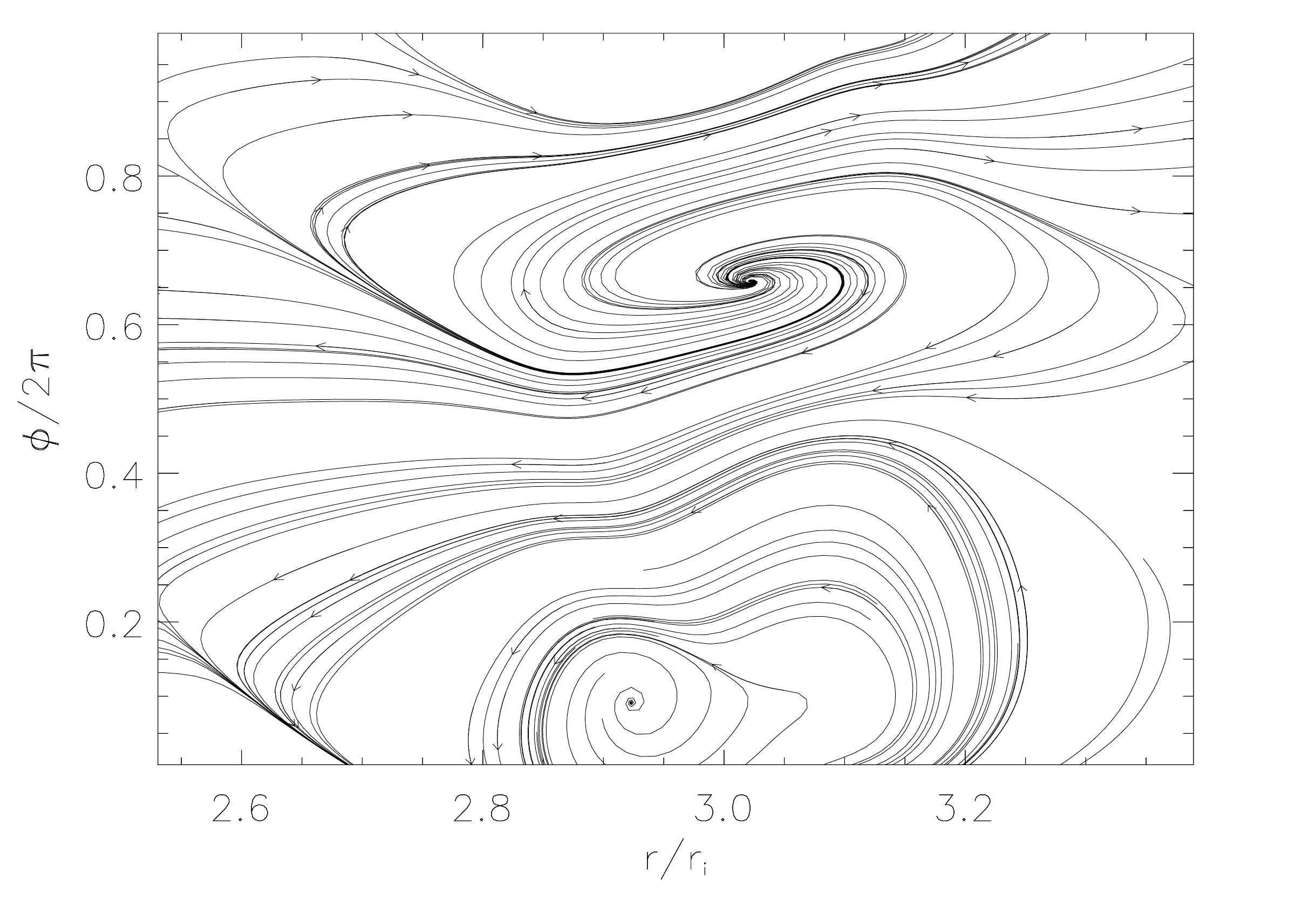}
   \includegraphics[width=8.8cm]{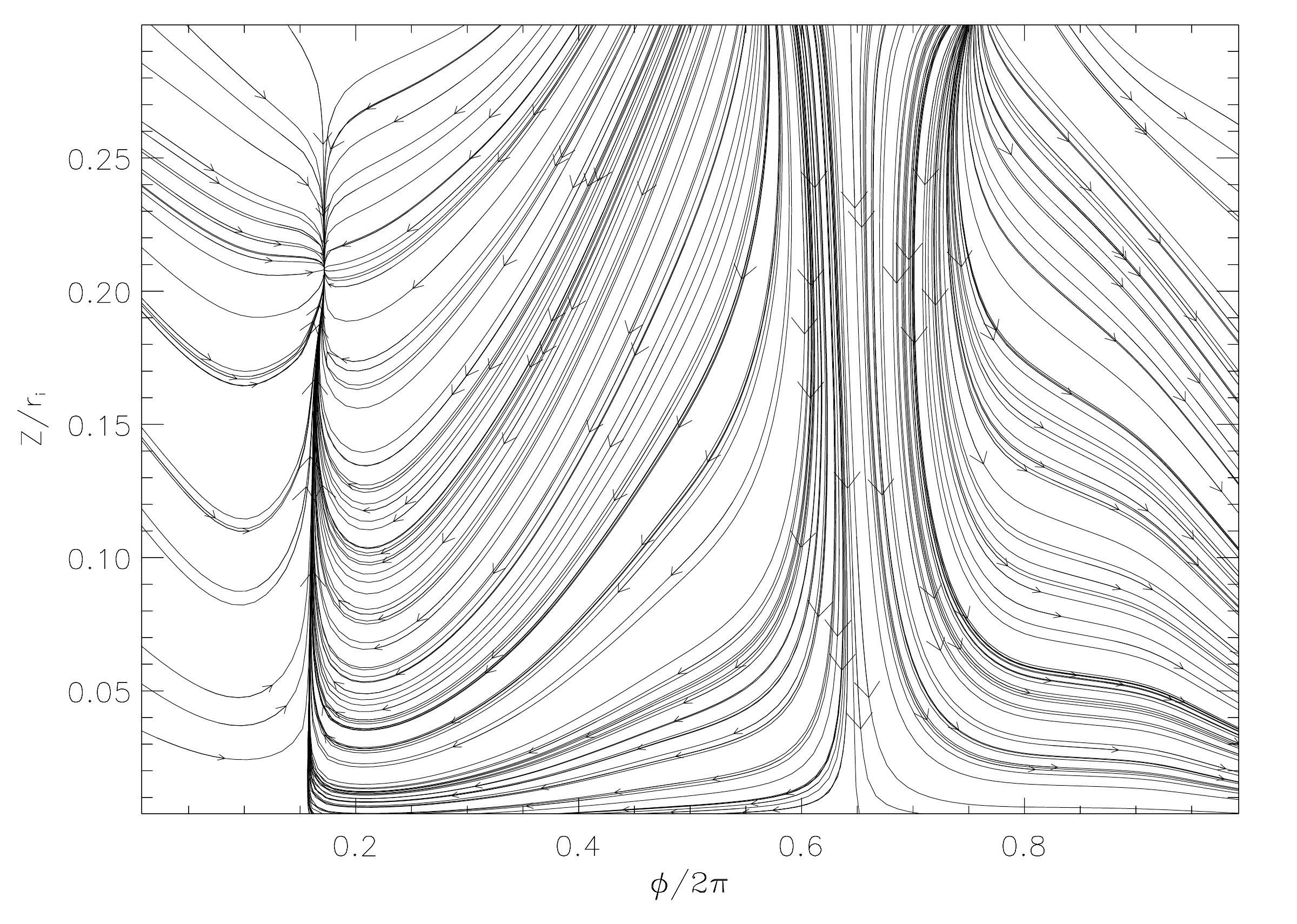}
\\
  \includegraphics[width=8.8cm]{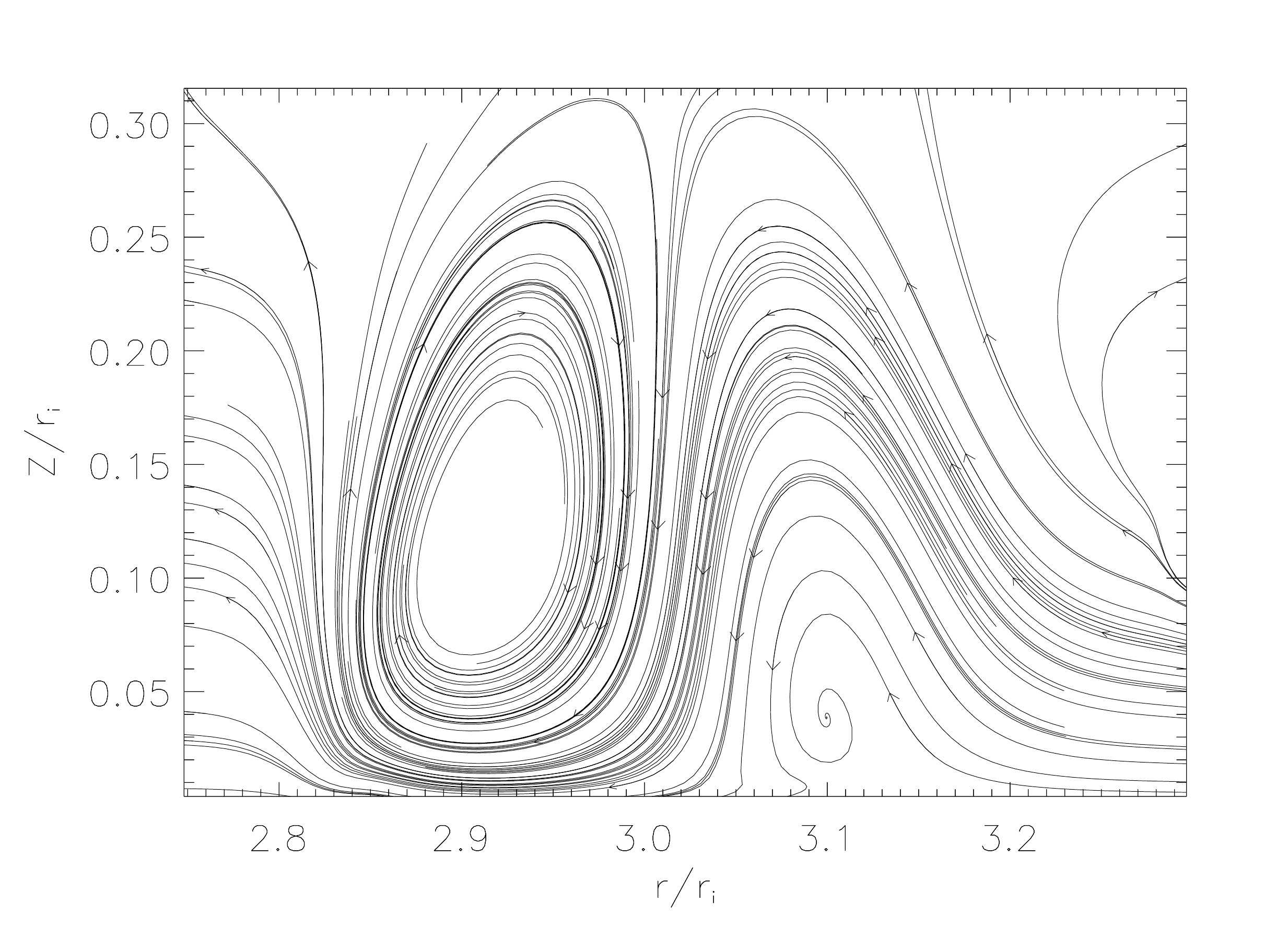}
   \includegraphics[width=8.8cm]{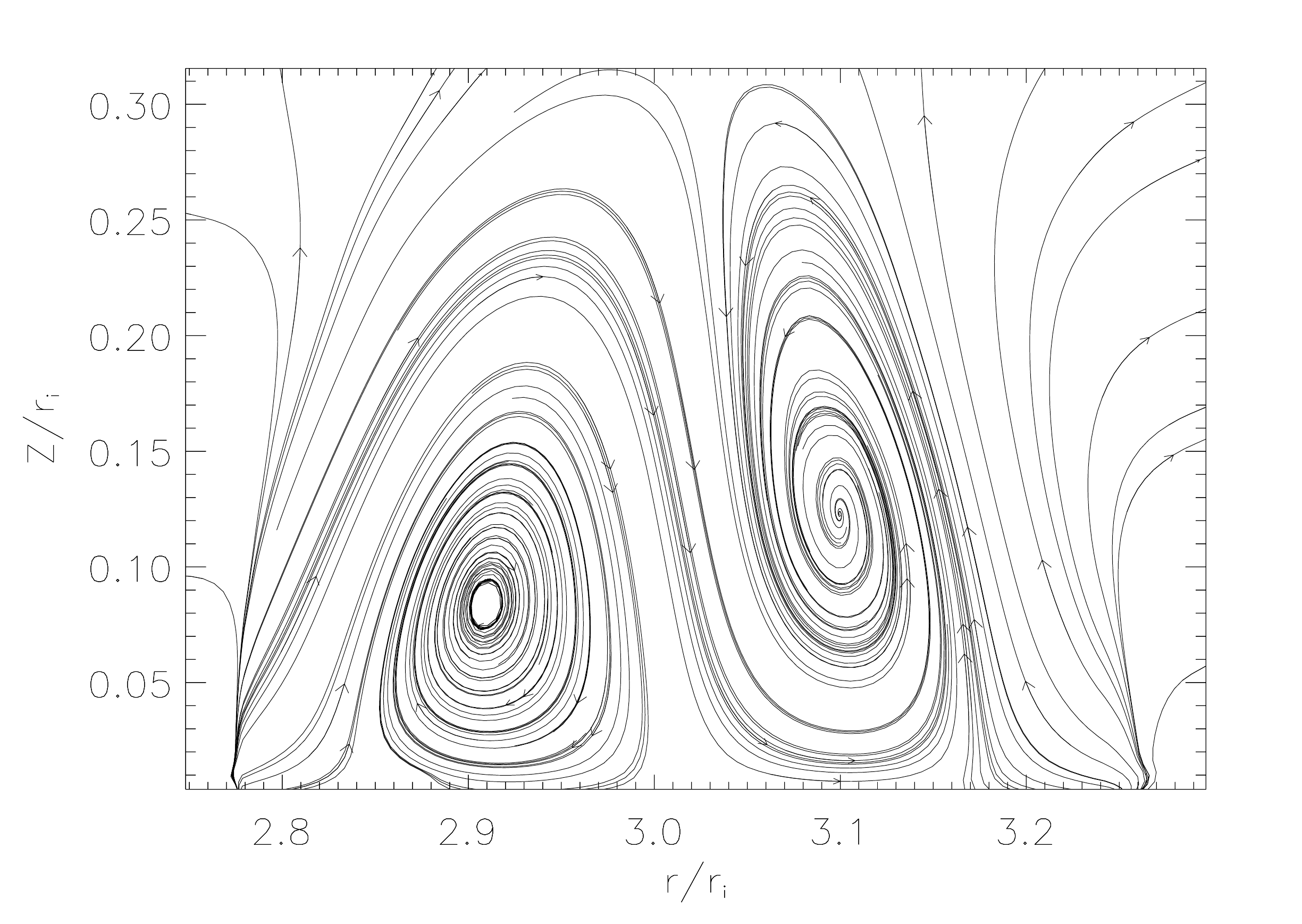}
      \caption{\emph{Top:} Velocity streamlines in a frame rotating with the disk, in the midplane of the disk (left) and in the vertical plane at $r=3r_i$ (right). \emph{Bottom:} Velocity streamlines in a vertical frame at $\phi/2\pi=0.64$ (left) and $0.66$ (right)}
         \label{FigCirculation}
   \end{figure*}
%

   \begin{figure}
   \centering
 \includegraphics[width=8.8cm]{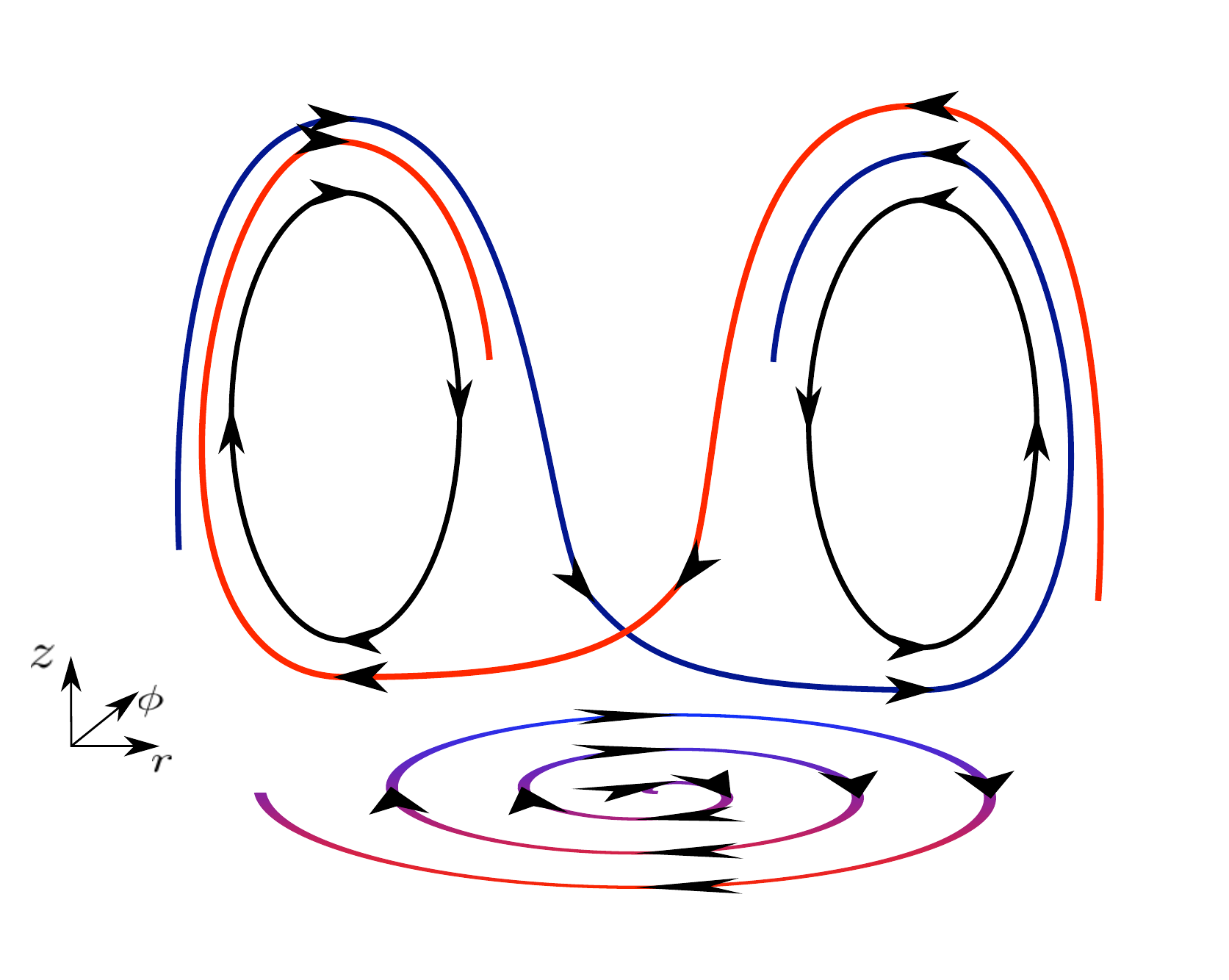}
      \caption{Schematic view of the circulation, showing the horizontal and vertical vortices in a sheared frame rotating at the local mean angular velocity.}
         \label{FigCirculationScheme}
   \end{figure}
%

In this section we consider the flow pattern generated by the instability. Since the stream in the $(r,\phi)$ plane is dominated by keplerian rotation, we consider the non axisymmetric part of the velocity, \ie we study the disk in a sheared frame rotating as the mean flow at each radius. Figure \ref{FigCirculation} does show in this plane the expected vortex structure, with both a cyclonic and anticyclonic feature since the $m=1$ structure generates both. As seen in figure \ref{fig:polaire} however, when combined with the keplerian flow only the anticyclonic component appears, forming a crescent-shaped feature, while the cyclonic one is at the x-point where the tips of the crescent join.

On the other hand figure \ref{FigCirculation} also shows unexpected new features: first, the plot in the ($\phi,\ z$) plane shows an downdraft at the center of the anticyclonic vortex, and a updraft at the center of the cyclonic one. Also, plots in the ($r,\ z$) plane show that strong convection rolls form on both sides of the vortex, in the anticlockwise direction for the outer one and clockwise direction for the inner one. Streamlines, at the outer edges of these horizontal and vertical vortices, connect them smoothly in a manner reminiscent of orbits near separatrices in hamiltonian dynamics. The main streamlines are schematized on figure \ref{FigCirculationScheme}. 

%
We have checked the robustness of these convection rolls by performing numerical simulations with the entire vertical structure of the disk, without assuming a vertical symmetry, all the other parameters kept unchanged. The vertical symmetry was broken by the initial perturbation added on the radial velocity, but resulted in similar flow patterns. Furthermore, when we applied only antisymmetric initial perturbations, we found that this delayed the development of the instability without changing its vertical structure. We conclude that the delay is due to the difficulty for antisymmetric initial perturbations to seed the unstable mode, and that the convection rolls are part of the linear structure of the instability. To confirm this diagnostic, we also checked that the maximal vertical velocity involved in these convection rolls, follows the linear growth of the instability as in figure \ref{FigGrowth}.

This strong vertical \new{structure} of the instability and of the vortex pattern could not be expected from the 2D results. We believe that the vertical vortices are resonantly excited where the local Doppler-shifted frequency, $\omega-m\Omega(r)$, is equal to the \gras{mean vertical oscillation frequency. The relation between this and the Br\"unt-V\"ais\"al\"a frequency, depending on the equation of state of the gas, will be discussed elsewhere. }
This would thus be the equivalent, in a fluid disk, of the \lq peanuts\rq ~structure found in observations and numerical simulations of barred spiral galaxies \citep{COM81,COM90,ANA08}. This is believed (although an explanation based on the firehose instability has also been proposed), to result from the resonant excitation of the vertical motion of stars  where $\omega-m\Omega(r)$ equals their vertical frequency of motion.

\section{\new{Discussion and outlooks}}

We have presented here the first fully three-dimensional, cylindrical numerical simulation showing a strong and persistent vortex in a differentially rotating disk. Its growth results from the Rossby Wave Instability, which has been invoked in various astrophysical contexts such as in protoplanetary disks or in the accretion disk of compact objects. 

The simulation essentially confirms the main properties of the instability, which forms a vortex extending across the vertically stratified structure of the disk. The vortex is localized at the extremum of vortensity and very efficiently acts to destroy this extremum, by permitting an exchange of angular momentum between regions located on either side of it. 

\gras{In our simulation we find no sign of the elliptical instability \citep{KER02,LES09} which has been claimed to destroy vortices. We note however that these works are done in the shearing sheet approximation \new{which neglects the equilibrium vorticity gradient and thus can} consider local small-scale modes whereas the RWI is a large-scale global instability. The same remark applies to the work of \cite{LAB09}.}

The most important new result found in our 3D simulation is the occurrence of strong vertical convection rolls, excited on both sides of the main, horizontal, vortex. Figure \ref{FigCirculation} shows that the vertical velocity in these rolls is comparable to the radial velocity involved in the Rossby vortex, while their growth shows that they are an inherent part of the structure of the mode, tightly connecting the horizontal and vertical circulations.

Future work should allow us to assess the importance of this convection in the astrophysical contexts where the RWI may act. In particular we expect them to play a very important role in the concentration of dust grains in protoplanetary disks, in addition to the mechanism of \cite{BAR95}, which will accumulate them at the center of the horizontal vortex. Indeed figure \ref{FigCirculation} also shows that, in the midplane of the disk, the flow lines rapidly spiral inward in the cyclonic vortex, and outward in the anticyclonic one. This goes together with the updraft at the cyclonic vortex and downdraft at the anticyclonic one. We note that dust grains will thus be rapidly transported with the gas, at low $z$, toward the center of the cyclonic vortex but that 
their weight should make it very difficult for them to be dragged in the updraft. \gras{The 3D structure of the vortex presented here will then have a direct impact on the accumulation of grains that will happen on an even lower timescale than in a 2D vortex.}
We also note however that, counter to both intuition and the effect discussed  by \cite{BAR95}, this would tend to accumulate the grains at the cyclonic rather than anticyclonic vortex. 

In order to explore this mechanism, which could be of primordial importance for the growth of planetesimals and planet formation, work should proceed in two directions: performing simulations using varied density profiles, including the one expected at the edge of the Dead Zone of protoplanetary disks \citep{VAR06}, and following the motion and growth of dust grains in the resulting vertical flows. An analytical description of the \gras{vertical} structure observed in the simulation is also needed.

\gras{Another direction for future work concerns the initial goal of this work, which was to analyze the 3D structure of MHD instabilities in magnetized disks.}
In accretion disks threaded by a vertical magnetic field, it has been shown that an Accretion-Ejection Instability (AEI) can occur \citep{TP99}, and explain the low-frequency Quasi-Periodic Oscillation of X-ray binaries \citep{ROD02, VAR02}. The instability grows by coupling magnetically-driven spiral density waves and a Rossby vortex. It has also been shown \citep{VT02} that this vortex can re-emit vertically, as Alfv\'en waves propagating to the corona along magnetic field lines, a substantial fraction of the accretion energy and angular momentum; it was also suggested that this could be a source for accretion-driven winds and jets. In this context we can expect that the convection rolls observed here would be replaced by resonantly excited slow magnetosonic waves. Since the main action of these waves is to move gas along magnetic field lines we can expect them to provide mass-load to the resulting jet, much as in MHD steady-state models of jets \citep{FEI95, CAF00} mass-loading occurs at the slow magnetosonic point and acceleration occurs in the region of the Alfv\'enic point. In order to study this prospect a fully 3D simulation that can handle both the geometry (a disk threaded by open poloidal magnetic field lines) and the magnitude (of the order of equipartition with the gas pressure) of the relevant magnetic field needs to be developed. 

\begin{acknowledgements}
This work was granted access to the HPC resources of IDRIS under the allocation 2009-i2009042125 made by GENCI (Grand Equipement National de Calcul Intensif).
\end{acknowledgements}

\bibliographystyle{aa}
\bibliography{RWI}

\begin{thebibliography}{37}
\expandafter\ifx\csname natexlab\endcsname\relax\def\natexlab#1{#1}\fi

\bibitem[{{Athanassoula}(2008)}]{ANA08}
{Athanassoula}, E. 2008, in IAU Symposium, Vol. 245, IAU Symposium, ed.
  {M.~Bureau, E.~Athanassoula, \& B.~Barbuy}, 93--102

\bibitem[{{Barge} \& {Sommeria}(1995)}]{BAR95}
{Barge}, P. \& {Sommeria}, J. 1995, \aap, 295, L1

\bibitem[{{Barranco} \& {Marcus}(2005)}]{BAR05}
{Barranco}, J.~A. \& {Marcus}, P.~S. 2005, \apj, 623, 1157

\bibitem[{{Casse} \& {Ferreira}(2000)}]{CAF00}
{Casse}, F. \& {Ferreira}, J. 2000, \aap, 353, 1115

\bibitem[{{Colella} \& {Woodward}(1984)}]{COL84}
{Colella}, P. \& {Woodward}, P.~R. 1984, Journal of Computational Physics, 54,
  174

\bibitem[{{Combes} {et~al.}(1990){Combes}, {Debbasch}, {Friedli}, \&
  {Pfenniger}}]{COM90}
{Combes}, F., {Debbasch}, F., {Friedli}, D., \& {Pfenniger}, D. 1990, \aap,
  233, 82

\bibitem[{{Combes} \& {Sanders}(1981)}]{COM81}
{Combes}, F. \& {Sanders}, R.~H. 1981, \aap, 96, 164

\bibitem[{{Falanga} {et~al.}(2007){Falanga}, {Melia}, {Tagger}, {Goldwurm}, \&
  {B{\'e}langer}}]{FAL07}
{Falanga}, M., {Melia}, F., {Tagger}, M., {Goldwurm}, A., \& {B{\'e}langer}, G.
  2007, \apjl, 662, L15

\bibitem[{{Ferreira} \& {Pelletier}(1995)}]{FEI95}
{Ferreira}, J. \& {Pelletier}, G. 1995, \aap, 295, 807

\bibitem[{{Goldreich} \& {Lynden-Bell}(1965{\natexlab{a}})}]{GOL65-1}
{Goldreich}, P. \& {Lynden-Bell}, D. 1965{\natexlab{a}}, \mnras, 130, 97

\bibitem[{{Goldreich} \& {Lynden-Bell}(1965{\natexlab{b}})}]{GOL65-2}
{Goldreich}, P. \& {Lynden-Bell}, D. 1965{\natexlab{b}}, \mnras, 130, 125

\bibitem[{{Johansen} {et~al.}(2004){Johansen}, {Andersen}, \&
  {Brandenburg}}]{JOH04}
{Johansen}, A., {Andersen}, A.~C., \& {Brandenburg}, A. 2004, \aap, 417, 361

\bibitem[{Johansen {et~al.}(2007)Johansen, Oishi, Low, Klahr, Henning, \&
  Youdin}]{JOH07}
Johansen, A., Oishi, J.~S., Low, M.-M.~M., {et~al.} 2007, Nature, Volume 448,
  Issue 7157, pp. 1022-1025 (2007).

\bibitem[{{Kerswell}(2002)}]{KER02}
{Kerswell}, R.~R. 2002, Annual Review of Fluid Mechanics, 34, 83

\bibitem[{{Lai} \& {Tsang}(2009)}]{LAI09}
{Lai}, D. \& {Tsang}, D. 2009, \mnras, 393, 979

\bibitem[{{Latter} \& {Balbus}(2009)}]{LAB09}
{Latter}, H.~N. \& {Balbus}, S.~A. 2009, \mnras, 399, 1058

\bibitem[{{Lesur} \& {Papaloizou}(2009)}]{LES09}
{Lesur}, G. \& {Papaloizou}, J.~C.~B. 2009, \aap, 498, 1

\bibitem[{{Li} {et~al.}(2001){Li}, {Colgate}, {Wendroff}, \& {Liska}}]{RWI3}
{Li}, H., {Colgate}, S.~A., {Wendroff}, B., \& {Liska}, R. 2001, \apj, 551, 874

\bibitem[{{Li} {et~al.}(2000){Li}, {Finn}, {Lovelace}, \& {Colgate}}]{RWI2}
{Li}, H., {Finn}, J.~M., {Lovelace}, R.~V.~E., \& {Colgate}, S.~A. 2000, \apj,
  533, 1023

\bibitem[{{Lovelace} \& {Hohlfeld}(1978)}]{LOV78}
{Lovelace}, R.~V.~E. \& {Hohlfeld}, R.~G. 1978, \apj, 221, 51

\bibitem[{{Lovelace} {et~al.}(1999){Lovelace}, {Li}, {Colgate}, \&
  {Nelson}}]{RWI1}
{Lovelace}, R.~V.~E., {Li}, H., {Colgate}, S.~A., \& {Nelson}, A.~F. 1999,
  \apj, 513, 805

\bibitem[{{Lyra} {et~al.}(2008){Lyra}, {Johansen}, {Klahr}, \&
  {Piskunov}}]{LYR08}
{Lyra}, W., {Johansen}, A., {Klahr}, H., \& {Piskunov}, N. 2008, \aap, 491, L41

\bibitem[{{Lyra} {et~al.}(2009){Lyra}, {Johansen}, {Zsom}, {Klahr}, \&
  {Piskunov}}]{LYR09}
{Lyra}, W., {Johansen}, A., {Zsom}, A., {Klahr}, H., \& {Piskunov}, N. 2009,
  \aap, 497, 869

\bibitem[{{Papaloizou} \& {Pringle}(1985)}]{PAP85}
{Papaloizou}, J.~C.~B. \& {Pringle}, J.~E. 1985, \mnras, 213, 799

\bibitem[{{Rodriguez} {et~al.}(2002){Rodriguez}, {Varni{\`e}re}, {Tagger}, \&
  {Durouchoux}}]{ROD02}
{Rodriguez}, J., {Varni{\`e}re}, P., {Tagger}, M., \& {Durouchoux}, P. 2002,
  \aap, 387, 487

\bibitem[{{Sellwood} \& {Kahn}(1991)}]{SEL91}
{Sellwood}, J.~A. \& {Kahn}, F.~D. 1991, \mnras, 250, 278

\bibitem[{{Tagger}(2001)}]{TAG01}
{Tagger}, M. 2001, \aap, 380, 750

\bibitem[{{Tagger} \& {Melia}(2006)}]{TAG06}
{Tagger}, M. \& {Melia}, F. 2006, \apjl, 636, L33

\bibitem[{{Tagger} \& {Pellat}(1999)}]{TP99}
{Tagger}, M. \& {Pellat}, R. 1999, \aap, 349, 1003

\bibitem[{{Tagger} \& {Varni{\`e}re}(2006)}]{TAV06}
{Tagger}, M. \& {Varni{\`e}re}, P. 2006, \apj, 652, 1457

\bibitem[{{T{\'o}th}(1996)}]{TOT96}
{T{\'o}th}, G. 1996, Astrophysical Letters Communications, 34, 245

\bibitem[{T\'{o}th \& Odstr\v{c}il(1996)}]{TOO96}
T\'{o}th, G. \& Odstr\v{c}il, D. 1996, J. Comput. Phys., 128, 82

\bibitem[{{Tsang} \& {Lai}(2009)}]{TSA09}
{Tsang}, D. \& {Lai}, D. 2009, \mnras, 400, 470

\bibitem[{{Varni{\`e}re} {et~al.}(2002){Varni{\`e}re}, {Rodriguez}, \&
  {Tagger}}]{VAR02}
{Varni{\`e}re}, P., {Rodriguez}, J., \& {Tagger}, M. 2002, \aap, 387, 497

\bibitem[{{Varni{\`e}re} \& {Tagger}(2002)}]{VT02}
{Varni{\`e}re}, P. \& {Tagger}, M. 2002, \aap, 394, 329

\bibitem[{{Varni{\`e}re} \& {Tagger}(2006)}]{VAR06}
{Varni{\`e}re}, P. \& {Tagger}, M. 2006, \aap, 446, L13

\bibitem[{{Yu} \& {Li}(2009)}]{YU09}
{Yu}, C. \& {Li}, H. 2009, \apj, 702, 75

\end{thebibliography}

\end{document}